\documentclass[aps,prl,twocolumn,floats,graphicx,epsfig,euscript,amsmath,amssymb]{revtex4}
\usepackage[dvips]{graphicx}
\usepackage{epsfig}
\DeclareMathAlphabet\EuScript{U}{eus}{m}{n} \SetMathAlphabet\EuScript{bold}{U}{eus}{b}{n}

\def\lapprox{\,\raise0.4ex\hbox{$<$}\kern-0.8em\lower0.7ex\hbox{$\sim$}\,}
\def\gapprox{\,\raise0.4ex\hbox{$>$}\kern-0.8em\lower0.7ex\hbox{$\sim$}\,}
\begin{document}

\bibliographystyle{prsty}

\title{Slow spin relaxation in a quantum Hall ferromagnet state}

\author{A.S. Zhuravlev$^{1}$, S. Dickmann$^{1}$, L.V. Kulik$^{1}$, I.V. Kukushkin${}^{1,2}$}
\affiliation{${}^1$Institute of Solid State Physics, RAS, Chernogolovka, 142432 Russia\\
${}^2$Max-Planck-Institut f\"ur Festk\"orperforschung, 1.$\,$Heisenbergstr.,
70569 Stuttgart, Germany}

\date{\today}

\begin{abstract}
{Electron spin relaxation in a spin-polarized quantum Hall state
is studied. Long spin relaxation
times that are at least an
order of magnitude longer than those
measured in previous experiments were observed and explained within
the spin-exciton relaxation formalism. Absence of any dependence of
the spin relaxation time on the electron temperature and on the
spin-exciton density, and specific dependence on the magnetic field
indicate the definite relaxation mechanism -- spin-exciton annihilation
mediated by spin-orbit coupling and smooth random potential.}

\noindent PACS numbers 73.43.Lp, 73.20.Mf, 73.21.Fg
\end{abstract}

\maketitle

The study of spin relaxation phenomena in two-dimensional electron systems (2DES) is a
broad and not yet well-understood field whose major benefactor is spintronics,
one of the most rapidly developing branches of applied physics and
technology \cite{Das Sarma04}. The dominant spin relaxation mechanism in a
2DES is typically associated with Rashba and Dresselhaus spin-orbit
(SO) couplings that occur due to the inversion asymmetry of the 2DES
confining potential and the host semiconductor. The Dyakonov-Perel
momentum-depending spin precession randomizes the spin orientation, thus
forcing spins to relax out of the initial direction
\cite{Dyakonov86,Leyland07}. A strong magnetic field normal to the 2DES
completely quantizes the energy spectrum, leading to a search for
alternative spin relaxation mechanisms. Surprisingly, despite the fact
that transport, optical, and magnetic phenomena in a 2DES in perpendicular
magnetic fields are well studied due to the integer and fractional quantum
Hall effects, spin relaxation physics are not well understood.
Experimentally observed spin relaxation times vary
from $\sim\!1\,$ns \cite{Alphenaar98} to $\sim\!10\,$ns \cite{Fukuoka,Nefyodov} and do
not provide a definite clue for identifying the major process governing the
relaxation of spins. Theoretical advances are more prominent; yet without a
proper experimental background these advances are characterized by a diversity of
studied relaxation mechanisms \cite{Frenkel,di04,di12} leading to a
wide range of calculated relaxation times. In this respect, the relevance
of some of the aforementioned calculations to the real 2DES is somewhat doubtful.
In addition, it is remarkable that all theoretically calculated times are much
longer than the times observed experimentally. {\em In the present
Letter, we report on spin relaxation measurements with relaxation
times reaching 150$\,$ns}, which strongly agree with theoretical
estimates for the ``disorder relaxation'' mechanism \cite{di12} (i.e., in which
the energy dissipation occurring simultaneously with the spin relaxation is
determined by a smooth random potential). Meanwhile, the observed
exponential time dependence for spin relaxation requires a special
explanation and therefore a more comprehensive theoretical analysis. In
particular, we first consider the importance of spin density spatial
fluctuations stimulated by the disorder.

The 2DES under quantum Hall conditions represents a quantum object
where Coulomb correlations radically modify the energy spectrum.
To date, the most elaborated spin-relaxation theory is the one
describing the quantum Hall ferromagnet at the electron filling factor
$\nu\!=\!1$ \cite{di04,di12}. This state is definitely a spin
dielectric where a deviation of the spin system from equilibrium
is treated as the appearance of {\em spin excitons} comprising
effective holes in the lowest spin sublevel of the zero Landau
level and electrons promoted to the next spin sublevel of the same
Landau level. The spin exciton with a nonzero momentum leads to
a reduction of the total spin and the spin projection along the
magnetic field of the 2DES by one; in turn, a zero
momentum spin exciton rotates the total spin of the 2DES in space
\cite{di04,di12}. Relaxation in both cases
is considered to represent an annihilation of spin excitons due to the SO
interaction. It is, however, a challenging task to verify this
theory because until now no experimental technique for creating
nonequilibrium systems with a considerable spin-exciton density
was available. Here, we report an optical technique suitable for this purpose. We chose time-resolved Rayleigh scattering \cite{Kulik12}as the methodology for monitoring the real-time
spin dynamics. The spin
relaxation times exceeded all currently known
experimental data by an order of magnitude.

\begin{figure}[htb!]
\advance\leftskip-0cm
\includegraphics[scale=.9,clip]{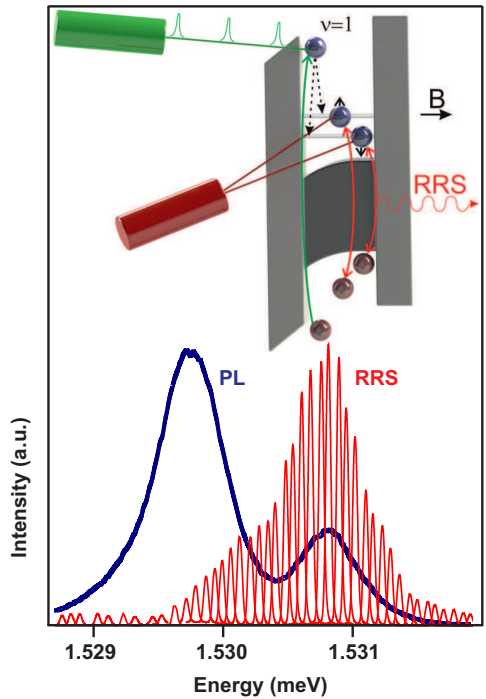}
\vspace{-2mm} \caption{\label{fig1} (top) A schematic diagram
of the experimental setup. (bottom) Photoluminescence (PL)
and resonant Rayleigh scattering spectra obtained 5~ms after the
heating laser pulse at a magnetic field of 11~T and a bath
temperature of 1.6~K.}\vspace{-8mm}
\end{figure}

Several GaAs/AlGaAs quantum wells of $18-20\,$nm width
2DES with an electron density in the range of $1.5-4\!\times\!10^{11}$cm$\!{}^{-2}$ were studied (dark mobility varying from $1$ to
$3\!\times\!10^6\,$cm$^2$/Vs). The experimental technique used
to produce a nonequilibrium spin system utilized short
($<\!\!1\,$ns) powerful laser pulses with the photon energy far
above the forbidden gap in GaAs (peak power was $10^4$J/cm$^2$),
see the illustration in Fig~1. By relaxing to the ground state, the
high-energy electrons heat the 2DES. The experiment has two
options: (i)~the characteristic time required to cool down the 2DES is
longer than the spin relaxation time (then the relaxation rate can
not be accurately measured) and (ii)~the spin relaxation time is
longer than the time required to cool down the 2DES. In the latter case,
the spin relaxation rate is measured directly.
This condition can be satisfied at magnetic fields
$B\!\geq\!9\,$T. The optically heated 2DES relaxes to a {\it
nonequilibrium} spin depolarized state. A continuous wave
radiation resonating with the optical transitions involving
electron and heavy-hole states in the zero Landau levels monitors
the 2DES spin by resonant Rayleigh scattering (RRS),
as described in Ref. \cite{Kulik12} (Fig.~1). The ratio of
the Rayleigh scattering cross sections for the two allowed optical
transitions (+1/2, -3/2) and (-1/2, +3/2) allows the
heating laser power to be tuned in such a way that the nonequilibrium 2DES
{proves to be} in an unpolarized state after cooling down.

The dynamics of the RRS signal in the $\nu\!=\!1$ quantum Hall ferromagnet
reveals nonmonotonic behavior. Within the first few
nanoseconds after the laser pulse, the RRS signal superimposes on the nonresonant photoluminescence induced by the pulse. The 2DES is then heated
during the first $10\,$ns. This process is accompanied by a complete loss
of the RRS intensity (Fig.~2). During the following $40\,$ns the RRS signal
returns to the initial value, thus indicating that the temperature of the 2DES
reaches the bath temperature. The RRS signal then again diminishes due to
the 2DES spin relaxation to the equilibrium state. Finally, the RRS signal is
saturated at a stable level, the magnitude of which depends on the bath temperature.
The stable level characterizes the equilibrium spin state of the 2DES
(Fig.~2). Measuring its magnitude at different temperatures allows us to determine
the temperature dependence of the 2DES equilibrium spin polarization. It
reproduces the well-known experimental results obtained with the nuclear magnetic
resonance technique (Fig. 2).

Subtracting the equilibrium RRS signal from the total RRS signal,
one obtains the spin relaxation rate as a function of the
nonequilibrium spin density (Fig.~2). We observed a number of
significant effects that reduced the number of possible mechanisms
accounting for the spin relaxation in 2DES. First, the relaxation time was not dependent on the electron temperature,
whereas the amount of relaxed electrons changed significantly.
[For example, at $4.2\,$K the relaxation is nearly exhausted
(Fig.~2).] Second, within the experimental accuracy, we found
no appreciable changes in the relaxation time versus the
nonequilibrium spin density (Fig.~2).

Another finding is the long spin relaxation time, which was more than
one order of magnitude longer than any known experimental results
\cite{Alphenaar98,Fukuoka,Nefyodov}. We believe that such a striking
discrepancy between our and other experimental data
was due to the different physical quantities measured. Our
technique directly aims at the spin relaxation from the upper to the lower
spin sublevel in a nonequilibrium strongly interacting 2DES
, whereas other experimental techniques, such as the
paramagnetic resonance and the Kerr rotation, focus on the spin
dephasing of a single or few noninteracting spin excitons.

\begin{figure}[htb!]
\vspace{-5mm}
\includegraphics[scale=1,clip]{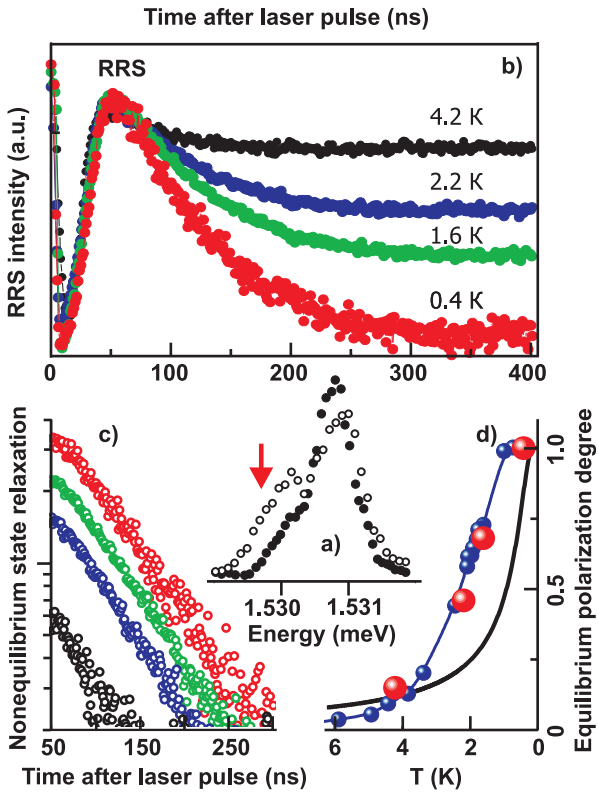}
\vspace{-5mm} \caption{\label{fig2} a) Envelopes for
nonequilibrium (open dots) and equilibrium (solid dots) RRS
spectra measured at B=11~T and T=1.6~K in 50~ns and 5~$\mu$s after
the heating laser pulse, respectively. The arrow indicates the
maximum of the PL line for the electron-hole transition associated
with the lowest spin Landau sublevel where the RRS signal dynamics
are measured. b) The RRS dynamics for the stable temperature points
of 0.4, 1.6, 2.2, and 4.2~K (the RRS signals at different
temperatures are normalized to give equal intensities 50~ns after
the heating laser pulse). c) The RRS signal intensity minus the
magnitude of the saturated RRS signal in logarithmic scale vs time
after the heating laser pulse at different bath temperatures
(point {colors} are the same as in b)). d) The
equilibrium spin polarization (large spheres) obtained using the
magnitudes of saturated RRS signals at different bath temperatures
(the equilibrium polarization at 0.4~K is taken as 1). For comparison, the
equilibrium spin polarization measured with the nuclear magnetic resonance
technique in
Ref. [\onlinecite{Barrett}] at a similar Zeeman energy (small
spheres connected by the thin line) is shown. The thick solid line
indicates single particle (without Coulomb correlations)
equilibrium spin polarization.}\vspace{-6mm}
\end{figure}

Spin excitons are true eigen states of the quantum Hall ferromagnet. A
comparative analysis of various temperature-independent relaxation channels
was recently reported \cite{di12}. We conclude that for $B\!<\!15\,$T,
the relaxation channel is determined by both the SO coupling providing the
spin nonconservation and the interaction with the smooth random potential
violating the momentum conservation. Indeed, a similar relaxation mechanism
was already theoretically discussed \cite {di04,di12}, but with one
significant difference -- smooth random potential was considered to be weak
and, therefore, producing no effect on the energy spectrum of spin excitons
or on their energy distribution. This was considered to be the cause of the
two-exciton scattering dissipation mechanism. As a result, this theoretical
model leads to a nonexponential relaxation law, $\sim 1/(1\!+\!t/\tau)$
\cite{di04,di12}, which thereby contradicts the observed exponential
relaxation. Some results, however, can be still utilized. Namely, let us consider a
2D domain with area $L^2$ where $L$ is much larger than the characteristic
correlation length $\Lambda$ of the correlator $K({\bf
r})\!=\!\langle\varphi({\bf r})\varphi(0)\rangle$ [$\varphi({\bf r})$
describes the external smooth random potential; $\langle\varphi({\bf
r})\rangle\,\equiv\,0$ is assumed].  In this case, the matrix element for the
transition from the two exciton state with wave vectors ${\bf q}_1$ and ${\bf
q}_2$ to a single exciton state with ${\bf q}'$ is $\left|{\cal M}({\bf q}_1,
{\bf q}_2,{\bf q}')\right|^2\!\!=\!{}4\pi\overline{K}({q}^*\!)
{{q^*}^2(u^2\!+\!v^2)}/{N_\phi^{2}}$, where ${\bf q}^*\!=\!{\bf q}_1\!+\!{\bf
q}_2\!-\!{\bf q}'$; $u\!=\!\sqrt{2}\alpha/l_B\hbar\omega_c$ and
$v\!=\!\sqrt{2}\beta/l_B\hbar\omega_c$ are the dimensionless SO constants
[$l_B$ -- magnetic length, $\omega_c$ -- cyclotron frequency, wave vectors
are measured in $1/l_B$ units]. We have taken the corresponding SO term in
the single electron Hamiltonian as ${\hat H}_{\rm SO}=\alpha\left(\hat{{\bf
q}}\times\hat{\mbox{\boldmath $\sigma$}} \right)_{\!z}\!+\!
\beta\left(\vphantom{\left(\hat{{\bf q}}\times\hat{\mbox{\boldmath $\sigma$}}
\right)}\hat{ q}_y\hat{\sigma}_y\!-\!{\hat
q}_x\hat{\sigma}_x\right),\:\mbox{where}\:\;\hat{{\bf q}}\!=\!-i{\mbox{\boldmath
$\nabla$}}\!+\!e{\bf A}/c\hbar$, $N_\phi\!=\!L^2/2\pi l_B^2$ is the number of
electron states in one spin sublevel, and
$\overline{K}({q})\!=\!\frac{1}{(2\pi)^2}\!\int\!K({\bf r})e^{-i{\bf q}{\bf
r}}d^2{r}$~is the correlator Fourier component. The annihilation is defined
by an elementary process in which two spin excitons with ${\bf q}_1$ and ${\bf
q}_2$ merge, and a new spin exciton arises with the energy ${\cal E}'$ equal
to the sum of energies ${\cal E}_1\!+\!{\cal E}_2$ for merging spin excitons.
The annihilation probability, if both excitons are within the domain
$L\!\times\!L$, is calculated with the standard formula $1/\tau({\bf
q}_1,{\bf q}_2)\!=\!\!\sum_{{\bf q}'}\!\frac{2\pi}{\hbar}\left|{\cal M}
  ({\bf q}_1,{\bf q}_2,{\bf
q}')\right|^2\!\delta\!\left({\cal E}_1\!+\!{\cal E}_2\!-\!{\cal E}'\right)$.

If the temperature is rather low, then to calculate the total
relaxation rate one has to know the quasi-equilibrium distribution
of ``cold'' spin-excitons determined not only by temperature but
also by the random potential (i.e., in the state where they are
cooled but not annihilated, because {the cooling processes
that do not involve the spin flip} are considered to occur much
faster). The problem of finding this distribution in real space
cannot be precisely solved, and we are only able to present a
comprehensive estimate. Now the exciton momentum is no longer a good
quantum number, and from the distribution in the conjugate
space one should return to the spin-exciton distribution in real
space characterized by the density $n({\bf r})$. From this point
of view, the values ${\bf q}_1$, ${\bf q}_2$ and ${\bf q}'$
entering the expressions above must be reexamined.

The following study is very similar to that presented recently
\cite{di13}. First we show how the random smooth potential
influences the spin-exciton energy. Let the random potential be
smooth (i.e., $l_B\!\ll\!\Lambda$). We consider a small domain
$l\times l$ ($l_B\!\ll\!l\!\ll\!\Lambda$ ) and use a
linear-gradient approximation: ${\hat
\varphi}\!=\!-(l_B/\hbar){\mbox{\boldmath
$\nabla$}}\!\varphi\!\times\!{\hat {\bf\cal P}}$ within this
domain, where ${\hat{\bf\cal P}}$ is a `generalized' momentum
operator of the magneto-exciton \cite{di13}. Within the
approximation of the first-order gradient expansion, the smooth
potential {does not change} exciton states and {conserves} the
quantum number ${\bf q}$. In addition, the linear potential {does not
contribute} to the value $\overline{K}({q}){q^2}$ [proportional to
the Fourier component of the correlator $\langle{\mbox{\boldmath
$\nabla$}}^2\!\!\varphi({\bf r})\varphi(0)\rangle$]. It does, however, determine
the electro-dipole part of the exciton energy so that
the latter is $\epsilon_{\rm Z}\!+\!q^2/2M_{\rm x}\!-\!{\bf d}{
\mbox{\boldmath $\nabla$}}\!\varphi$, where ${\bf d}\!=\!l_B({\bf
q}\!\times\!{\hat z})$ is the dipole moment, ($\epsilon_{\rm
Z}\!=\!g\mu_BB$ is the Zeeman gap, and $1/M_{\rm x}$ is the
exciton inverse mass; $q$ is in the $1/l_B$ units). After cooling
down, but before annihilation, the exciton in the vicinity of
point ${\bf r}$ acquires wave vector ${\bf q}_m\!\equiv\!M_{\rm
x}l_B{\hat z}\! \times\!{\mbox{\boldmath $\nabla$}}\!\varphi({\bf
r})$ and ``gets stuck'' in the smooth random potential with the
exciton energy ${\cal E}_m\!=\!\epsilon_{\rm Z}-q_m^2/2M_{\rm x}$.
The inverse `local' relaxation time $1/\tau({\bf q}_1,{\bf q}_2)$
is calculated with substitution ${\bf q}_1\!=\!{\bf q}_2\!=\!{\bf
q}_m({\bf r})$, and as a result, if the correlator is Gaussian,
$K(r)\!=\!\Delta^2e^{-r^2/\Lambda^2}\!$, one obtains $\tau({\bf
r})\!=\!N_\phi {\cal T}({\bf r})$, where here and in the
expressions above one has to change from $N_\phi\!=\!L^2/2\pi
l_B^2$ to $N_\phi\!=\!l^2/2\pi l_B^2$. Thus, one finds\vspace{-1mm}
\begin{eqnarray}\label{tau}
  \frac{1}{{\cal T}({\bf r})}=\frac{M_{\rm x}\Delta^2\Lambda^2(u^2\!+\!v^2)}{\hbar l_B^2}\quad\qquad{}\qquad\qquad\nonumber\\\times
  \int_{q'_{\rm min}}^{q'_{\rm max}}\!\!\frac{\exp{\![-q^2(q')\Lambda^2/4]}\, q^2(q')\,q'dq'}{\sqrt{q_m^2q'^2-(q'^2/2+q_m^2-\epsilon_{\rm Z}M_{\rm x})^2}}\,. \vspace{-1mm}
\end{eqnarray}
Here $q_m^2({\bf r})\!=\!(\!M_{\rm x}l_B\!{\mbox{\boldmath $\nabla$}}\!\varphi)^2$, $q'_{\rm min}\!=\!|\sqrt{2M_{\bf x}{\cal E}_m}\!-\!q_m|$, $q'_{\rm
max}\!=\!\sqrt{2M_{\bf x}{\cal E}_m}\!+\!q_m$, and $q^2(q')\!=\!4\epsilon_{\rm
Z}M_{\rm x}\!-\!q'^2$ \cite{foot}.
\vspace{-0mm}
\begin{figure}[htb!]
\vspace{-5mm}
\includegraphics[scale=0.95,clip]{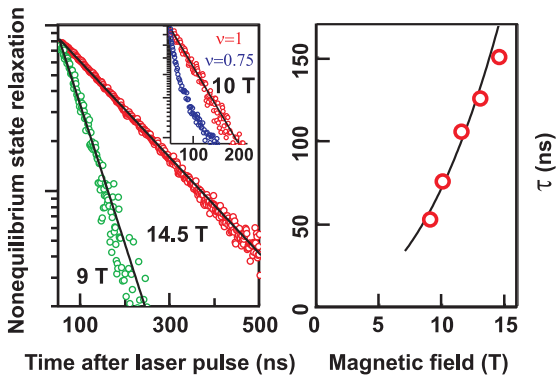}
\vspace{-4mm}\caption{\label{fig3}\vspace{-0mm}  (left) The RRS signal intensity minus the
magnitude of the saturated RRS signal in the logarithmic scale
(points) vs time after the heating laser pulse measured for B=9
and 14.5~T. Solid lines being exponential fits to the experimental
data. Purple rings on the inset represent the RRS signal corresponding
to spin-relaxation dynamic at $\nu\!=\!0.75$ and $B\!=\!10\,$T. The
relaxation is actively non-exponential in the case.
For comparison, we also show the $\nu\!=\!1$ data at the same magnetic
field. (right) The relaxation time constant for the quantum Hall
ferromagnet vs magnetic field (open dots). Solid line is the
theoretical fit (value of ${\EuScript T}$) to the experimental
data obtained with the parameter set indicated in the
text.}\vspace{-3mm}
\end{figure}

The experimentally measurable quantity -- the total relaxation
rate -- is calculated by multiplying the probability $1/\tau({\bf
r})$ by the probability of finding two spin excitons within the
$l\!\times\!l$ domain, $\times n^2({\bf r})l^4$, and then by
summation over all domains. The latter means substitution
$l^2\!\to\!d{\bf r}$ and integration over ${\bf r}$:\vspace{-2mm}
 \begin{equation}
 -dN_{\rm x}/dt\!=\!2\pi l_B^2\!\int\!n^2({\bf r})/{\cal T}({\bf r})d{\bf r}.\vspace{-2mm}
\end{equation}
Thus, we have a complex problem of finding the density
distribution $n({\bf r})$ and calculating the integral in Eq. (2). A quasi-equilibrium distribution $n({\bf r})$ is determined by
fast transition processes preceding the spin exciton annihilation:
cooling due to the electron-phonon coupling and simultaneous
drifting in the presence of a smooth random field. This field
influencing the exciton energy ${\cal E}_m$ is actually not
$\varphi$ but $({\mbox{\boldmath $\nabla$}}\!\varphi)^2\!$,
representing a non-Gaussian case.

We obtain the required estimate in two steps. First, let us change
from ${\cal T}({\bf r})$ to the averaged value ${\Tilde{\cal T}}$
by substituting for $q_m^2$ in Eq. (1) the mean quantity $\langle
q_m^2\rangle$ calculated with the averaging procedure $\langle
q_m^2\rangle/(M_{\rm x}l_B)^2\!=\!\langle({\mbox{\boldmath
$\nabla$}}\!\varphi)^2\rangle\!\equiv\!\!2(\Delta/\Lambda)^2$, which is
valid for the Gaussian potential $\varphi$. Second, for a spatially
fluctuating density $n({\bf r})\!\equiv\!N_{\rm
x}/\!L^2\!+\!\delta n({\bf r})$, we estimate the integral
$\int\!\!n^2({\bf r})d{\bf r}\!\equiv\!N_{\rm
x}^2/\!L^2\!+\!\int[\delta n({\bf r})]^2d{\bf r}$, where the last
term is proportional to the spatial correlator $\langle \delta
n({\bf r})\delta n(0)\rangle$ at ${\bf r}\!=\!0$. Should spin
excitons form an ideal gas, this correlator would correspond to
so-called white noise and would be equal to $\delta({\bf
r})N_{\rm x}/L^2$ \cite{ll91}. In our case, the correlations are
mostly determined by spatial fluctuations of the field
$({\mbox{\boldmath $\nabla$}}\!\varphi)^2$: if it is energetically
favorable to find a spin exciton at a point ${\bf r}_0$, then the
density should be over the mean value, $\delta n({\bf r}_0)\!>\!0$
in the neighborhood $|{\bf r}\!-\!{\bf r}_0|\!\lesssim\!\Lambda'$.
For an estimate, the $\delta({\bf r})$ function is replaced
with a `hat'-function $e^{-r^2/{\Lambda'}^2}\!\!/\pi{\Lambda'}^2$.
Intuitively, the estimate for $\Lambda'$ is
$\Lambda'\!\simeq\!\Lambda/2$. As a result, we come to
$\int[\delta n({\bf r})]^2\!d{\bf r}\!\sim\!4N_{\rm
x}/\pi{\Lambda}^2$.

Thus, there are two contributions to the relaxation rate (2). The
first contribution is a quadratic in $n_{\rm x}\!=\!2\pi l_B^2N_{\rm x}/L^2$
and is determined by the mean excitonic density, whereas the
second contribution originates from the part of the fluctuations that
is linear in $n_{\rm x}$:\vspace{-2mm}
\begin{equation}
-dn_{\rm x}/dt\sim n^2_{\rm x}/{\Tilde {\cal T}}+8l_B^2n_{\rm x}/{\Tilde {\cal T}}\Lambda^2.\vspace{-1mm}
\end{equation}
Our theoretical approach is based on neglecting the
inter-excitonic correlations. That is, $n_{\rm x}$ is assumed to
be small, and the second term in Eq. (3) is thus dominant.
Experimentally, the observed relaxation is well-exponential in
time, even starting from the initial value $n_{\rm x}(0)\simeq
0.5$. So, semi-empirically, we conclude that the characteristic
relaxation time is given by ${\EuScript T}\!=\!{\tilde {\cal
T}}\Lambda^2/8l_B^2$. The material parameters can
be evaluated using available data for similar quantum wells
\cite{Larionov} and by slightly varying the poorly known quantities:
$\Delta$ and $\Lambda$, around their experimentally estimated
values. The magnitude of $1/M_{\rm x}$ was borrowed from a recent
experiment \cite{Kulik}. In so doing, and to compare the theory with our experimental
measurements, one could, for example, consider the following specific
parameters quite reasonable for our experimental conditions:
$\alpha\!=\!0.25\,$nm$\cdot$meV, $\beta\!=\!0.12\,$nm$\cdot$meV,
$\epsilon_{\rm Z}\!=\!0.02B\,$meV, $1/M_{\rm
x}\!=\!0.87B^{1/2}$meV ($B$ in Teslas), $\Delta\!=\!1.25\,$meV,
and $\Lambda\!=\!32\,$nm. As a result, one obtains a specific
dependence ${\EuScript T}(B)$ that nicely describes the experimental
data (Fig.~3). Finally, note that the real temperature at which
the employed `zero-temperature' approach should be valid is
estimated to be $T\!<\!{\langle q_m^2\rangle}/M_{\rm
x}\!\sim\!M_{\rm x}(l_B\Delta/\Lambda)^2\!\simeq\!1\,$K.

Finally as a basis for a discussion we briefly concern the situation of a softer quantum Hall ferromagnet. In case the filling deviates from one, the spin polarization diminishes (see, e.g., an analysis in Ref.\cite{pl09} ), and drastic rearrangement of the ground state -- emergence of a skyrmionic texture has to result in radical changes of excitations' picture. In particular, it is known that additional spin-flip modes are observed\cite{ga08}, and even only due this fact spin-relaxation physics becomes quite different compared to the $\nu\!=\!\!1$ case. Our intuitive opinion of this interesting but hardly studied relaxation problem enables us to think that the phase volume relevant to spin relaxation processes should grow if $\nu$ deviates from one -- the relaxation thus occurs faster. Our experimental technique allows, e.g., to observe spin-relaxation dynamics at $\nu\!=\!0.75\,\,$(see the inset on Fig.$\,\,$3).

In conclusion, we emphasize the importance of the presented results.
Up to now, experimental physics have been focused on indirectly investigating
the spin dephasing time of noninteracting spin excitons.
In our study, we successfully prepared a nonequilibrium spin
system that allows us to {\em perform real-time monitoring of the spin
dynamics}. On the basis of these data and theoretical analysis, we
describe a scenario for the relaxation and conclude that, in fact,
there is no discrepancy between the experimental results and
theoretical notions based on the concept of spin-exciton
annihilation -- the observed relaxation can be explained within
the framework of the SO--smooth-disorder relaxation mechanism, and actually
looks like effective inter--spin-exciton scattering.
In addition, the long relaxation times, $100-150\,$ns, provide a
chance to study a possible condensate state of nonequilibrium
spin excitons.

The authors acknowledge the Russian Fund of Basic Research for
support. S.D. is grateful to Serge Florens for valuable discussions
concerning the fundamental problems of averaging under the conditions
of a non-Gaussian smooth random potential. 

\end{document}